\documentclass[sigconf]{acmart}

\usepackage{multirow}
\usepackage{booktabs}
\usepackage{subfigure}
\usepackage{graphicx}
\usepackage{amsmath}

\usepackage{tikz}
\newcommand{\mybox}[1]{\tikz[baseline=(MeNode.base)]{\node[rounded corners=0.75mm, fill=gray!20](MeNode){#1};}}
\usepackage{longfbox}
\usepackage{stfloats}

\AtBeginDocument{%
  }

\acmPrice{15.00}
\acmISBN{978-1-4503-XXXX-X/18/06}

\begin{document}

\title{Prompt-Based Zero- and Few-Shot Node Classification: A Multimodal Approach}


\author{Yuexin Li}
\email{yuexinli@u.nus.edu}
\orcid{0000-0001-8525-2853}
\affiliation{%
  \institution{National University of Singapore}
  \country{Singapore}
  \postcode{119077}
}

\author{Bryan Hooi}
\email{dcsbhk@nus.edu.sg}
\affiliation{%
  \institution{National University of Singapore}
  \country{Singapore}
  \postcode{119077}
}

\renewcommand{\shortauthors}{Li et al.}

\begin{abstract}
Multimodal data empowers machine learning models to better understand the world from various perspectives. 
In this work, we study the combination of \emph{text and graph} modalities, a challenging but understudied combination which is prevalent across multiple settings including citation networks, social media, and the web. 
We focus on the popular task of node classification using limited labels; in particular, under the zero- and few-shot scenarios. 
In contrast to the standard pipeline which feeds standard precomputed (e.g., bag-of-words) text features into a graph neural network, we propose \textbf{T}ext-\textbf{A}nd-\textbf{G}raph (TAG) learning, a more deeply multimodal approach that integrates the raw texts and graph topology into the model design, and can effectively learn from limited supervised signals without any meta-learning procedure. 
TAG is a two-stage model with
(1) a prompt- and graph-based module which generates prior logits that can be directly used for zero-shot node classification, and
(2) a trainable module that further calibrates these prior logits in a few-shot manner.
Experiments on two node classification datasets show that TAG outperforms all the baselines by a large margin in both zero- and few-shot settings.
\footnote{Work in progress}
\end{abstract}





\maketitle

\section{Introduction}
Much of the data we encounter in practical applications is \emph{multimodal}, comprising information of different modalities, such as image and text, video and audio, and so on. Driven by this, \emph{multimodal machine learning}, which aims to process and combine information from multiple modalities, has become a rapidly growing field. Prominent applications include video captioning \cite{video-caption, video-caption-2, video-caption-3, video-caption-4} which combines video and audio data, large scale self-supervised learning using image and text data \cite{self-supervised-image-and-text, self-supervised-image-and-text-2, self-supervised-image-and-text-3}, emotion recognition \cite{emotion-recognition, emotion-recognition-2, emotion-recognition-3}, activity recognition \cite{activity-recognition, activity-recognition-2, activity-recognition-3, activity-recognition-4}, and many others.

In this work, we focus on the combination of \emph{text and graph} modalities. While this combination is much less well-studied in prior work, data that combines these two modes are still commonplace particularly in web-related settings: for example, citation networks \cite{citation-networks}, knowledge graphs \cite{knowledge-graph-dbpedia}, misinformation detection on social media \cite{misinformation}, linked text databases such as Wikipedia \cite{wikipedia}, and the web itself as a massive set of hyperlinked pages. 

As a concrete task, we focus on one of the most popular graph-related tasks, namely node classification using limited labels, commonly referred to as semi-supervised node classification \cite{gcn}. We first consider the zero-shot case, where no labeled nodes are available, but we have the name of each class. We then consider the few-shot case, where we have a small number of labels from each class. 
\begin{figure}
    \centering
    \includegraphics[scale=0.5]{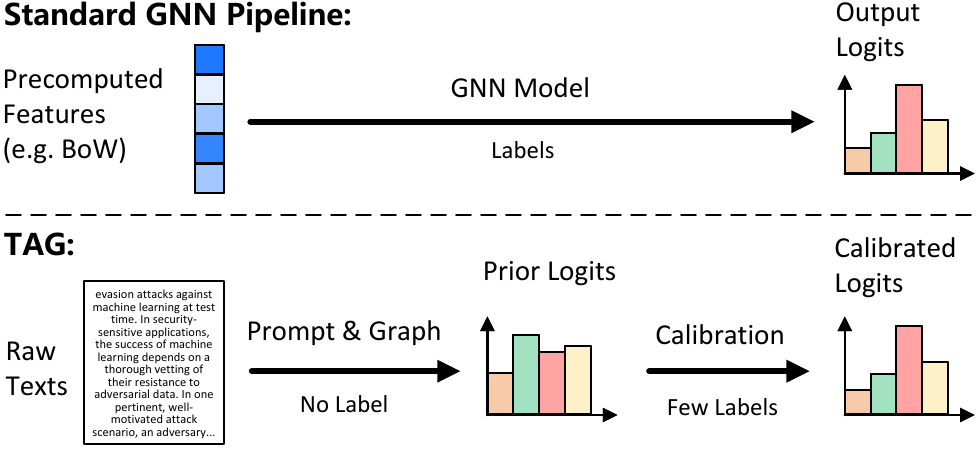}
    \caption{A comparison between standard GNN pipeline and our multimodal approach TAG for node classification.}
    \label{fig:intro}
\end{figure}

The predominant approach for node classification relies on graph neural networks (GNNs). In the standard pipeline, we are given a graph (e.g., a citation graph) with node attributes being standard precomputed text features, such as bag-of-words (BoW) features. Indeed, the popular node classification datasets such as \textsc{ogbn-arxiv}, \textsc{cora}, \textsc{pubmed}, and \textsc{citeseer} almost always rely on such precomputed text features as node attributes, so that almost all existing node classification methods are developed on top of these features as input.

In contrast, in our work, we show the benefits of our fully multimodal approach, termed \textbf{T}ext-\textbf{A}nd-\textbf{G}raph (TAG) learning, where the text information on each node is used not just in the form of precomputed features, but integrated in the design of the method itself. Figure \ref{fig:intro} provides a comparison between standard GNN pipeline and TAG. In particular, under the zero and few-shot settings, we develop an approach combining pretrained prompt-based language models with a graph-based enhancement module, leading to large performance gains: for example, improving accuracy by 20.0\% in the zero-shot setting and 21.5\% in the 3-shot setting on \textsc{ogbn-arxiv}, and 20.3\% in the zero-shot setting and 8.1\% in the 3-shot setting on \textsc{ogbn-products}, compared to the best performing baseline. We argue that  deeply incorporating multimodality in this way opens new avenues for model design: in particular, our approach relies on the raw text data to perform zero-shot learning via prompts, which could not be done if we were relying on precomputed text features.

\begin{figure}
    \centering
    \includegraphics[scale=0.5]{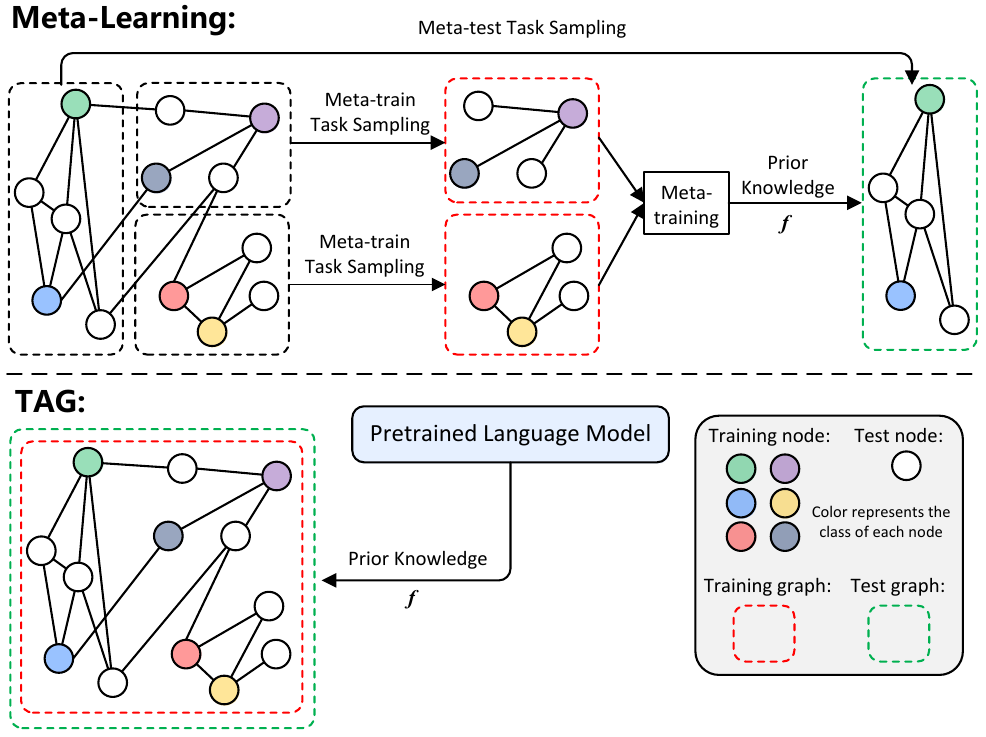}
    \caption{A comparison between meta-learning-based few-shot node classification approaches and TAG.}
    \label{fig:intro2}
\end{figure}

Besides its strong performance, TAG also has significant benefits in terms of simplicity and ease-of-use compared to existing methods for zero- and few-shot learning on graphs, which primarily rely on meta-learning. As shown in Figure \ref{fig:intro2}, the differences between TAG and meta-learning-based approaches are mainly three-fold. 
\textbf{1) The set of classes}: Meta-learning-based approaches require to meta-train
their node classification models on many node classification tasks. This is usually achieved by sacrificing a set of classes in a dataset to perform meta-training, while the remaining classes can be used for meta-testing. In contrast, TAG does not need any meta-training and can perform inference on all the classes in a dataset.
\textbf{2) The graphs}: Meta-learning relies on other graphs to form meta-training tasks (mostly by partitioning the graph in a dataset), while TAG can perform inference directly on target graph without the need of any additional graphs. 
\textbf{3) The ways to acquire prior knowledge}: Instead of meta-training the node classification model on many node classification tasks to acquire prior knowledge, TAG obtains its prior knowledge directly by prompting pretrained language models, and thus effectively skips the meta-training stage. 
In other words, the introduction of prompt in TAG avoids the complex meta-learning procedure, leading to much more simplicity for zero- and few-shot node classification tasks.

In summary, our contributions are:
\begin{itemize}
    \item \textbf{Multimodal text and graph approach}: in contrast to existing node classification approaches which rely on precomputed text features as input to graph neural networks, we show the benefits of combining raw text with graphs in a more deeply multimodal approach.
    \item \textbf{Zero- and few-shot learning without meta-learning}: our novel approach combines pretrained prompt-based language models with a graph-based enhancement module, and can perform zero and few-shot learning without the need for meta-learning. 
    \item \textbf{Large performance gains}: our approach improves accuracy by 20.0\% to 20.3\% in the zero-shot setting and 8.1\% to 21.5\% in the 3-shot setting compared to the best performing baseline.
\end{itemize}

\section{Related Work}

\subsection{Graph Neural Networks}
Recent years have witnessed remarkable progress on the development of GNNs, as they are capable of learning rich patterns in graph-structured data. 
Popular models such as GCN \cite{gcn} and GAT \cite{gat} rely on message passing to update the embeddings of each node. To address the large scale challenge, minibatching GNNs are also developed by sampling neighbors \cite{graphsage} or subgraphs \cite{graphsaint} to learn node embeddings in an inductive way. Other works also tackle the scalability issue by proposing simplified methods like SGC \cite{sgc} and LINKX \cite{linkx}. 

In additional to design new GNN architectures, there are also works on improving the node features. For example, instead of using bag-of-word features, GIANT-XRT \cite{giant-xrt} feeds the raw text of each node into a language model to obtain new node features via self-supervised extreme multi-label classification, where the graph topology is used to construct the labels of each node. These new features help existing GNNs achieve better node classification performance than the bag-of-word features do. 

These aforementioned approaches are designed for node classification, but do not consider how to handle the zero- and few-shot settings. We will next discuss how GNNs can be equipped with zero- and few-shot learning ability.

\subsection{Zero- and Few-shot Learning on Graphs}\label{sec:related_work} The general few-shot learning pipeline leverages prior knowledge to generalize to new tasks with very limited supervision \cite{few_shot_survey}. MAML \cite{maml} is the most popular framework used in graph few-shot learning tasks \cite{gfsl-survey}. The prior knowledge gained by MAML is the model parameter initialization, which is obtained by pretraining the model on a variety of similar tasks (i.e., meta-training). Such parameter initialization enables the model to adapt quickly to a new task with only a small number of training samples (i.e., meta-testing).

For few-shot node classification, methods that adapt MAML to GNNs include meta-training on similar node classification tasks \cite{meta-gnn, gpn, gfl} and local subgraphs \cite{g-meta}, such that the GNNs are well initialized for meta-test tasks. Meta-PN \cite{meta-pn} combines meta-learning and label propagation in graphs to augment labeled data to learn a GNN. 

In contrast, there are very few works on zero-shot node classification. DGPN \cite{dgpn} performs zero-shot knowledge transfer by using class semantic descriptions to transfer knowledge from a set of seen classes to a disjoint set of unseen classes, which is similar to the meta-learning setting. A few other methods utilize graphs as a tool to perform zero-shot learning in other domains such as image classification \cite{rgen, apnet, cge}, rather than being designed for graph data as input.

Our approach is fundamentally different from the above methods, because it does not require training on data from a disjoint set of classes for prior knowledge acquisition, as is required by meta-learning methods and DGPN \cite{dgpn}. Instead, our prior knowledge is computed by the prior logits generation module with the help of pretrained language models. This reduces computational cost and avoids the need for suitable data drawn from a disjoint set of classes as the testing data, which may not be available in practice.

\subsection{Prompt-based Learning}

Prompt-based learning provides text instructions to the pretrained language models (PLMs) to allow PLMs to adapt faster to downstream tasks such as text classification \cite{prompt_classification} and question answering \cite{prompt_qa}. PET \cite{pet} and ADAPET \cite{adapet} realize zero- and few-shot text classification by augmenting labeled data with prompt model. These augmented data in turn serves as sufficient training samples for standard supervised learning. LM-BFF \cite{lm-bff} applies prompt-based finetuning with automatic selected prompt templates, which shows better performance than vanilla finetuning. As such pipelines offer auxiliary supervised signals to finetune the PLMs, they are also categorized as prompt tuning.

Our approach also utilizes the power of prompt models, but does not require PLM finetuning, making our approach significantly easier to use and more efficient, as modern PLMs can be massive in terms of number of parameters. We directly obtain logits by feeding prompts into a PLM, which is then processed by a graph-based module. We show that such utilization of text and graph modalities leads to large performance gains.

\section{Methodology}
\subsection{Problem Statement}
A graph $\mathcal{G}$ is defined as $\mathcal{G}=(\mathcal{V}, \mathbf{A})$, where $\mathcal{V}$ and $\mathbf{A}\in\mathbb{R}^{|\mathcal{V}|\times|\mathcal{V}|}$ represent the set of nodes and the adjacency matrix, respectively. Each node $v\in\mathcal{V}$ belongs to one of the $C$ classes in the task, and is associated with its input text $s$ together with its true label text $l\in\mathcal{L}$ (e.g. `Artificial Intelligence' or `Databases'), where $\mathcal{L}$ is the set of all label texts and $|\mathcal{L}|=C$. 
We further define an injective function $\rho: \mathcal{L}\rightarrow \mathbb{N}^+$ that maps a label text to its class index.

The node set $\mathcal{V}$ is partitioned into the training set $\mathcal{V}_{\textrm{train}}$ and the test set $\mathcal{V}_{\textrm{test}}$. Specifically, for the zero-shot setting, the whole node set is directly used as the test set (i.e.,$\mathcal{V}_{\textrm{test}}=\mathcal{V}$) and $\mathcal{V}_{\textrm{train}}$ is not available. For the few-shot setting, we consider the commonly studied $N$-way $K$-shot classification task, where $K$ labeled samples are given for each of the $N$ classes. In this work, we set $N=C$, thus we have $|\mathcal{V}_{\textrm{train}}|=CK$ and $|\mathcal{V}_{\textrm{test}}|=|\mathcal{V}|-CK$.
 
Given $\mathcal{V}_{\textrm{train}}$, our goal is to obtain a node classification model $f$ which can accurately classify all the nodes in $\mathcal{V}_{\textrm{test}}$ in both zero- and few-shot settings.

\subsection{Overview}
\begin{figure*}[t]
    \centering
    \includegraphics[scale=0.6]{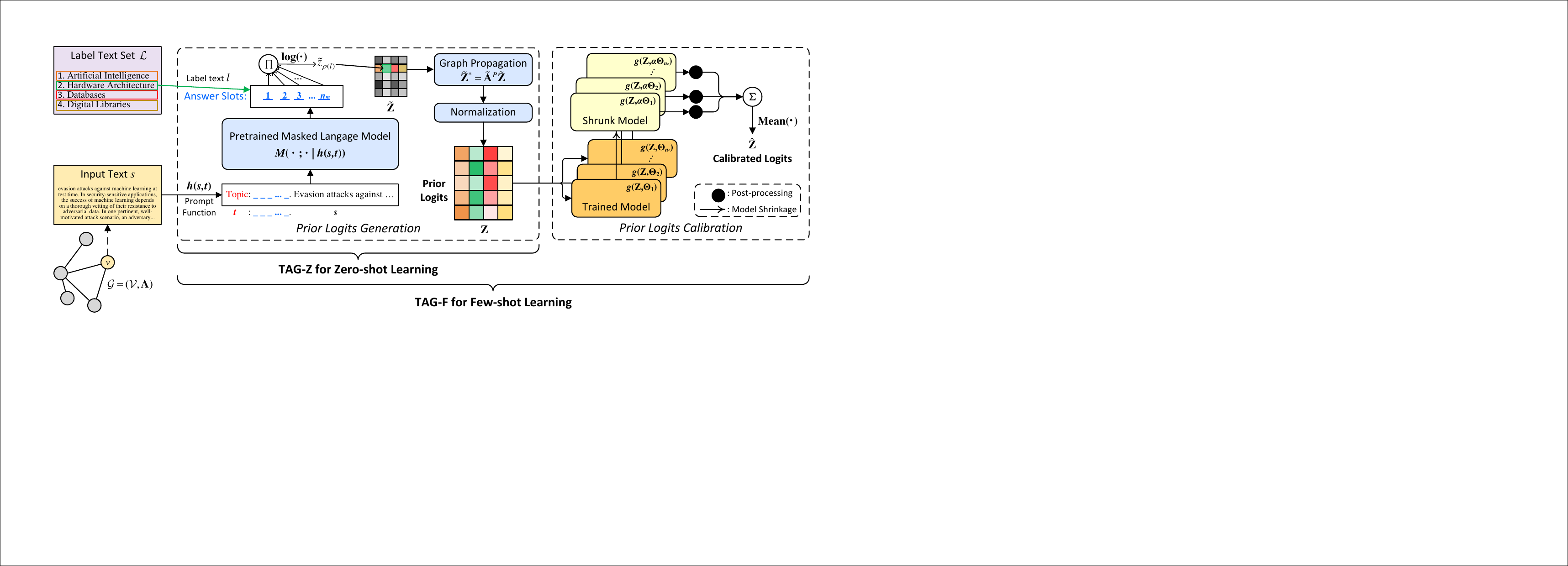}
    \caption{An overview of our multimodal node classification approach TAG.}
    \label{fig:TAG}
\end{figure*}
We propose \textbf{T}ext-\textbf{A}nd-\textbf{G}raph (TAG) learning---a multimodal node classification approach that learns from text and graph data in a more direct way. It comprises the following two main modules:
\begin{enumerate}
    \item \textbf{Prior Logits Generation (PLG)}: uses raw texts and the adjacency matrix to generate prior logits for each node via a prompt- and graph-based approach;
    \item \textbf{Prior Logits Calibration (PLC)}: learns to calibrate the prior logits and output the corrected logits for prediction.
\end{enumerate}
As we show in Figure \ref{fig:TAG}, our zero-shot learning approach, TAG-Z, consists of only the PLG module. Our few-shot learning approach, TAG-F, is the integration of both the PLG and PLC modules. 

\subsection{Prior Logits Generation}
Our PLG module aims to generate the \emph{prior logits} $\mathbf{Z}\in\mathbb{R}^{|\mathcal{V}|\times|\mathcal{L}|}$, which is equivalent to a distribution over classes for each node. Given the input and label texts, we leverage the language modelling ability of large-scale pretrained masked language models (PMLMs) to obtain the prior logits. This is realized by feeding a \emph{prompt}---a simple modification of the original text into a text completion task, into a PMLM. By doing so, we utilize the text modality in a more direct way, instead of simply using precomputed features such as bag-of-words. Subsequently, since neighbouring nodes tend to be similar or dependent, it is also reasonable to improve the quality of these logits with the help of the graph topology. Hence, we introduce a prompt-based module and a graph-based module in PLG.

\subsubsection{Prompt}
The prompt function $h(s,t)$ we use in this work has the following template form:
\begin{equation}
    h(s, t)=\mybox{$t. \_\_\_\_. s$}
\end{equation}
which modifies the original input text $s$ into a combination of an instruction text $t$, multiple mask tokens \mybox{${\_\_\_\_}$} as answer slots, and $s$ itself. The number of mask tokens in the answer slots, denoted $n_m$, is equal to the maximum token length among all the labels in $\mathcal{L}$. 


Denote $M$ as a PMLM. Each label text $l\in\mathcal{L}$ is a sequence of tokens $(l_1, l_2, ..., l_{n_l})$, where $n_l\le n_m$ is the number of tokens in $l$. We can then obtain a score $\Tilde{z}_{\rho(l)}$, which represents the unnormalized log probability of an input text $s$ belonging to class $\rho(l)$:
\begin{equation}
\begin{split}
    \Tilde{z}_{\rho(l)} = \log\prod_{i=1}^{n_l} M(i;l_i|h(s,t)) = \sum_{i=1}^{n_l}\log M(i;l_i|h(s,t))
\end{split}
\label{eq:prompt}
\end{equation}
where $M(i;l_i|h(s,t))$ outputs the probability score that token $l_i$ appears in the $i$-th masked slot given the prompt $h(s,t)$ as the input of $M$.

By repeating Eq. (\ref{eq:prompt}) for each label text $l\in\mathcal{L}$ for an input text $s$ of a node $v\in\mathcal{V}$, we can obtain a score vector $\Tilde{\mathbf{z}}=[\Tilde{z}_1, \Tilde{z}_2, ..., \Tilde{z}_{|\mathcal{L}|}]\in\mathbb{R}^{|\mathcal{L}|}$. We can further obtain a score matrix $\Tilde{\mathbf{Z}}\in\mathbb{R}^{|\mathcal{V}|\times|\mathcal{L}|}$ by concatenating the score vectors of all the nodes in $\mathcal{V}$ together.

\subsubsection{Graph Propagation}
A graph propagation step is used to update a node's score vector by averaging the score vectors of that node and all its first-order neighbor nodes. We start with constructing a modified adjacency matrix $\mathbf{A}'=\mathbf{A} + \mathbf{I}$ by converting directed edges into undirected edges and adding self-loops. Let $\mathbf{D}$ denote a diagonal matrix whose $i$-th diagonal entry is equal to the sum of all the entries in the $i$-th row of $\mathbf{A}^{'}$. The updated score matrix $\Tilde{\mathbf{Z}}^*$ is calculated by performing $P$ steps of graph propagation with the normalized adjacency matrix $\Tilde{\mathbf{A}}=\mathbf{D}^{-1}\mathbf{A}'$ to $\Tilde{\mathbf{Z}}$:
\begin{equation}
    \Tilde{\mathbf{Z}}^{*} = \Tilde{\mathbf{A}}^{P} \Tilde{\mathbf{Z}}
\end{equation}

\subsubsection{Normalization}
The product operation in Eq. (\ref{eq:prompt}) can potentially lead to scale bias of the score values. For example, the probability score of a one-token label could be much larger than that of a ten-token label regardless of what the input text $s$ is. This can result in a highly biased class distribution where the class of the one-token label always dominates. 

We mitigate this problem by normalizing the logits for each class over all the samples. 
Let $\Tilde{\mathbf{Z}}^{*}_{:,j}\in\mathbb{R}^{|\mathcal{V}|}$ be the $j$-th column in $\Tilde{\mathbf{Z}}^{*}$, then the normalization operation is represented as:
\begin{equation}
    \mathbf{Z}_{:,j} = \frac{\Tilde{\mathbf{Z}}^{*}_{:,j} - \mu(\Tilde{\mathbf{Z}}^{*}_{:,j})}{\sigma(\Tilde{\mathbf{Z}}^{*}_{:,j})}
    \label{eq:normaliaztion}
\end{equation}
where $\mu(\cdot)$ and $\sigma(\cdot)$ are the mean and standard deviation operators, respectively.

\subsubsection{Zero-shot Classification} 
After normalization, we obtain the prior logits $\mathbf{Z}=[\mathbf{Z}_{:,1}, \mathbf{Z}_{:,2}, ..., \mathbf{Z}_{:,|\mathcal{L}|}]$, which can be directly used for zero-shot inference. For instance, the predicted class $\hat{y}_{i}$ of node $v_i$ can be inferred from the $i$-th row vector of $\mathbf{Z}$:
\begin{equation}
    \hat{y}_{i} = \mathop{\arg\max}_{j=1,2,..., |\mathcal{L}|} \mathbf{Z}_{i, j}
\end{equation}

\subsection{Prior Logits Calibration}
Given the prior logits $\mathbf{Z}$, we next refine these logits with the help of a few labeled samples. To do this, we use our PLC module, which is designed to calibrate in a stable and controlled manner even under the few-shot setting, avoiding overfitting. PLC generates the \emph{calibrated logits} $\hat{\mathbf{Z}}\in\mathbb{R}^{|\mathcal{V}|\times|\mathcal{L}|}$ by introducing a trainable model and a special ensemble of these models as the calibration module.

\subsubsection{Trainable Model}
For simplicity, we use multi-layer perceptron (MLP) with $Q$ layers as part of the trainable model $g(\cdot, \Theta)$, where $\Theta$ is the set of trainable parameters in an MLP. Specifically, each layer involves a linear transformation sublayer, a batch-norm sublayer, and a ReLU activation in order. Given the output $\mathbf{X}^{(q-1)}$ from the ($q-1$)-th layer, the forward computations of the $q$-th layer are:
\begin{align}
    \mathbf{U}^{(q)} &= \mathbf{X}^{(q-1)}\mathbf{W}+\mathbf{b}^{(q-1)} \\
    \mathbf{V}^{(q)} &= {\rm{BatchNorm}} (\mathbf{U}^{(q)}) \\
    \mathbf{X}^{(q)} &= {\rm{ReLU}}(\mathbf{V}^{(q)})
\end{align}
where $\mathbf{X}^{(0)}=\mathbf{Z}$. We add $\mathbf{Z}$ as an identity connection to the output of the MLP, and the output of the trainable model is represented as:
\begin{equation}
    g(\mathbf{Z}, \Theta) = \mathbf{U}^{(Q)} + \mathbf{Z}
    \label{eq:id}
\end{equation}
Note that the identity connection is similar to the residual connection in ResNet \cite{resnet}, but it is instead a scalar that does not contain any gradient information during training. Such formulation, together with the Shrinkage Ensemble (discussed in section \ref{sec:shrinkage}), makes full use of the prior logits as prior knowledge to foster the few-shot learning process. 

\subsubsection{Optimization}
We optimize $g(\mathbf{Z}, \Theta)$ by minimizing a combination of two loss functions. One is the cross-entropy loss $\mathcal{J}_{\textrm{CE}}$ computed over $\mathcal{V}_{\textrm{train}}$, and the other is the entropy loss $\mathcal{J}_{\textrm{H}}$ computed over $\mathcal{V}$. Minimizing $\mathcal{J}_{\textrm{H}}$ helps the model to generate more informative logits by incorporating unlabelled data, which mitigates the lack of labels. We use a coefficient $\lambda\in(0,1)$ to $\mathcal{J}_{\textrm{H}}$ to balance the effect of these two loss functions. Denoting the ground truth class vector as $\mathbf{y}\in\mathbb{R}^{|\mathcal{V}|}$, the combined loss is:
\begin{equation}
    \mathcal{J} = \mathcal{J}_{\textrm{CE}}(g(\mathbf{Z}, \Theta), \mathbf{y}) + \lambda \mathcal{J}_{\textrm{H}}(g(\mathbf{Z}, \Theta))
\end{equation}
where $\mathcal{J}_{\textrm{H}}(g(\mathbf{Z}, \Theta))= \frac{1}{|\mathcal{V}|}\sum_{i=1}^{|\mathcal{V}|}\sum_{j=1}^{|\mathcal{L}|}\mathbf{P}_{i,j}\log(\mathbf{P}_{i,j})$ and $\mathbf{P}$ is a probability score matrix obtained from $g(\mathbf{Z}, \Theta)$ after applying the softmax operation.

\subsubsection{Shrinkage Ensemble}\label{sec:shrinkage}
The few-shot setting leads to overfitting and high random variability of the trained model between different random seeds, especially in the absence of a validation set for model selection (e.g., early stopping). 

Hence, in this section we introduce the novel idea of mitigating such random variability using \emph{post-processing via shrinkage toward the prior}. Specifically, after training $\Theta$, we multiply all parameters in $\Theta$ by a \emph{shrinkage coefficient} $\alpha\in(0,1)$; note that as $\alpha$ gets close to $0$, the term $\mathbf{U}^{(Q)}$ approaches $0$, and thus $g(\mathbf{Z},\alpha\Theta)$ approaches the prior $\mathbf{Z}$. Shrinkage toward the prior, as well as averaging over an ensemble, both help to counteract the adverse impact of random variability, considering that the prior logits $\mathbf{Z}$ already have fairly good classification performance. 


To do this, we first obtain $\mathbf{S}:=g(\mathbf{Z}, \alpha\Theta)$---the output of the shrunk model. 
We next apply a \emph{scaling} operation to post-process $\mathbf{S}$.
Given the $j$-th column $\mathbf{S}_{:,j}\in\mathbb{R}^{|\mathcal{V}|}$ of $\mathbf{S}$, the scaling operation is:
\begin{align}
    \hat{\mathbf{S}}_{:,j} &= \frac{\mathbf{S}_{:,j}}{\epsilon + \sigma(\mathbf{S}_{:,j})} 
\end{align}
where $\epsilon=10^{-8}$ is used to avoid division by zero. We thus obtain the post-processed logits $\hat{\mathbf{S}}=[\hat{\mathbf{S}}_{:,1}, \hat{\mathbf{S}}_{:,2}, ..., \hat{\mathbf{S}}_{:,|\mathcal{L}|}]$ by concatenating all the columns together. 

Denoting $n_e$ as the ensemble size, we train $n_e$ separate model copies in the above way, to get $\hat{\mathbf{S}}^{(i)}$ for $i=1, \cdots, n_e$ as the post-processed logits of the $i$-th shrunk model $g(\mathbf{Z}, \alpha\Theta^{(i)})$. Then, the calibrated logits $\hat{\mathbf{Z}}$ is the mean of these $n_e$ post-processed logits:
\begin{equation}
    \hat{\mathbf{Z}} = \frac{1}{n_e}\sum_{i=1}^{n_e} \hat{\mathbf{S}}^{(i)}
\end{equation}

\subsection{Efficiency}
The most significant efficiency gain of TAG comes from our more direct utilization of the PLMs, which are prompts, for prior knowledge acquisition. We will elaborate this on the way we handle text and graph modalities and how such way benefits zero- and few-shot node classification. 

On the one hand, we handle the text and graph modalities without training any language model. The only existing approach that makes deep use of text and graph modalities, GIANT-XRT \cite{giant-xrt}, requires to train a language model with extreme multi-label classification to extract node features. The training of language models (usually have a huge number of parameters) introduces much more training cost, while the node features it generates cannot be directly used for zero-shot inference. In contrast, TAG does not need to train any language models, due to the use of prompts and graph propagation to obtain prior logits. The prior logits (a distribution over all the classes for each node), as the prior knowledge we obtain for zero- and few-shot node classification, can be directly used for zero-shot inference and foster few-shot learning. 

On the other hand, the use of prompts avoids the need to perform further meta-training or data augmentation. The current meta-learning-based methods, such as Meta-GNN \cite{meta-gnn}, GPN \cite{gpn}, and G-Meta \cite{g-meta} have to obtain a model parameter initialization (as their prior knowledge) by meta-training on similar node classification tasks. Meta-PN \cite{meta-pn} also needs to meta-learn feature-label transformer with additional unlabelled nodes that are less likely to obtain in practice. TAG, instead, does not require any meta-training or data augmentation in advance and can directly perform inference on all the classes in a dataset with the help of prior logits obtained from prompts and graphs. This makes TAG an easy-to-use zero- and few-shot node classification model.

\section{Experiments}
We aim to answer the following research questions through the experimental studies:
\begin{itemize}
    \item \textbf{RQ1 (Effectiveness)}: Can our proposed approach TAG excel at the zero- and few-shot node classification tasks? 
    \item \textbf{RQ2 (Ablation studies)}: How do the prompts and each component of TAG contribute to its overall performance?
    \item \textbf{RQ3 (Effects of Hyperparameters)}: What are the effects of different graph propagation steps $P$, ensemble size $n_e$, and shrinkage coefficient $\alpha$ on the performance of TAG?
\end{itemize}

\subsection{Datasets}
As our focus is on the text-and-graph multimodal setting, only datasets providing raw texts on each node can be considered. Based on this criteria, two publicly available node classification datasets, \textsc{ogbn-arxiv} and \textsc{ogbn-products} \cite{ogb}, are used to evaluate our proposed TAG approach. Table \ref{tab:datasets} provides an overview of these two datasets.
\begin{table}[htbp]
\renewcommand{\arraystretch}{.9}
  \caption{Overview of the datasets used.}
  \label{tab:datasets}
  \centering
  \begin{tabular}{cccc}
      \toprule
      \textbf{Dataset} & \textbf{\#Nodes} & \textbf{\#Edges} & \textbf{\#Classes}\\
      \midrule
      \textsc{ogbn-arxiv}    & 169343    & 1166243   & 40 \\
      \textsc{ogbn-products} & 2449029   & 61859140  & 47 \\
      \bottomrule
  \end{tabular}
\end{table}

\subsection{Baselines}

For the zero-shot setting, we compare TAG-Z with two zero-shot natural language inference baselines: 1) \textbf{Sentence Embedding Similarity (SES)}; and 2) \textbf{Prompt} \cite{pet}. For SES, we generate the sentence embedding (SE) \cite{sentence-bert} for the input text of each node and the label texts, after which the class with the highest cosine similarity score between the input text embedding and the label text embedding will be the zero-shot prediction of that node. For Prompt, they share the same prompt functions as those in TAG. Each of the two baselines is integrated with four pretrained language models, which are \textbf{BERT} \cite{bert}, \textbf{RoBERTa} \cite{roberta}, \textbf{DistilBERT} \cite{distilbert} and \textbf{DistilRoBERTa}. Besides, we also include \textbf{Rand-Guess} (performing random prediction), \textbf{Rand-GCN} and \textbf{Rand-GAT} (randomly initialized GNNs taking RoBERTa SE as input but without training) as the naive baselines.

For the few-shot setting, we compare TAG-F with seven baselines: \textbf{MLP}, \textbf{GCN} \cite{gcn}, \textbf{GAT} \cite{gat}, \textbf{SGC} \cite{sgc}, \textbf{GCNII} \cite{gcn2}, \textbf{UniMP} \cite{unimp}, \textbf{RevGNN} \cite{revgnn}. We consider two choices of input features for each of these baselines: 1) the features provided by OGB \cite{ogb} (denoted as OGB$^{\dagger}$), and 2) the SE of the input text of each node computed by RoBERTa-base.

We do not consider the few-shot methods mentioned in section \ref{sec:related_work} (e.g. DGPN for zero-shot, and Meta-GNN for few-shot) as baselines because they require meta-training or pretraining on other node classification tasks. This usually involves splitting the set of classes in a dataset into two disjoint class sets---a known class set for meta-training, and a novel class set for meta-testing, as in many existing works \cite{dgpn, meta-gnn, gpn, gfl, g-meta, meta-pn}, but this is incompatible with our setting where such meta-training data is not assumed to be available.
Moreover, it is infeasible to perform meta-training on one dataset (e.g. \textsc{ogbn-arxiv}) and meta-testing on another dataset (e.g. \textsc{ogbn-products}) due to the inconsistent number and nature of the classes in both datasets. For example, if these models are pretrained on $N$-way $K$-shot tasks, they have to be tested on tasks with the same $N$. This restriction, on the other hand, highlights the ease-of-use of our TAG model, which does not require meta-training, flexibly enabling inference on all the classes on any text-and-graph dataset. 

\subsection{Experimental Setup}
We use RoBERTa-base as the PMLM for TAG, unless specifically specified. The prompt functions $h(s,t)$ in TAG are $h$($s$, `Topic') for \textsc{ogbn-arxiv} and $h$($s$, `Product\ Category') for \textsc{ogbn-products}. The number of graph propagation steps $P$ is 10 for both datasets. For TAG-F, we set the number of layers $Q$ to 3, the hidden channels of $g(\cdot, \Theta)$ to 16, the entropy loss coefficient $\lambda$ to 0.3, the shrinkage coefficient $\alpha$ to 0.9, and the ensemble size $n_e$ to 5. We adopt the Adam optimizer to train TAG-F with learning rate $10^{-2}$ and a fixed number of epochs 50. 

All the few-shot baselines have 128 hidden channels in total for fair comparison; more details on model architecture are shown in Appendix section \ref{sec:app_setup}. They also share the same number of layers and are trained with the same optimizer, learning rate, and number of epochs, as TAG-F. For the large-scale \textsc{ogbn-products} dataset that poses scalability issues for inductive GNNs, we apply minibatch training to GCN, GAT, GCNII, UniMP, and RevGNN by sampling 10 neighbors at each layer, while using full-batch for inference. 

We perform the few-shot experiments 5 times for all methods, where each time differs in the sampling of $\mathcal{V}_{\textrm{train}}$ and $\mathcal{V}_{\textrm{test}}$. We use the mean test accuracy (ACC) and the weighted F1-scores (F1) over these 5 repetitions for performance comparison, as they are the most commonly used metrics in node classification. 

All the models are implemented in PyTorch \cite{pytorch} version 1.10.2 via PyTorch Geometric \cite{pyg} and Transformers libraries. We perform model training and inference on a Ubuntu server with Intel(R) Xeon(R) Gold 6226R CPU @ 2.90GHz and 4 NVIDIA RTX 3090 24GB graphics cards.

\subsection{RQ1: Effectiveness}
\begin{table}[htbp]
\renewcommand{\arraystretch}{.9}
    \caption{Zero-shot test ACC (\%) and F1 (\%) of TAG-Z and the baselines on two datasets.}
    \label{tab:zero-shot-main}
    \centering
    \begin{tabular}{llccccc}
    \toprule
    \multirow{3}{*}{\textbf{Method}} & \multirow{3}{*}{\textbf{PMLM}} & \multicolumn{2}{c}{\textsc{arxiv}} & \multicolumn{2}{c}{\textsc{products}} \\
    \cmidrule(l){3-4} \cmidrule(l){5-6}
     & & \textbf{ACC} & \textbf{F1} & \textbf{ACC} & \textbf{F1} \\
    \midrule
    Rand-Guess   & NA      &  2.50 &  3.22 &  2.13 &  3.03 \\ 
    Rand-GCN     & RoBERTa &  3.04 &  0.43 &  1.38 &  0.21 \\ 
    Rand-GAT     & RoBERTa &  1.13 &  0.32 &  2.56 &  0.26 \\ 
    \midrule
        \multirow{4}{*}{SES}
        & BERT              &  7.43 &  5.95 &  3.04 &  0.94 \\ 
        & RoBERTa           & 13.08 &  9.51 &  6.25 &  3.73 \\ 
        & DistilBERT        &  9.47 &  8.89 &  3.34 &  1.23 \\ 
        & DistilRoBERTa     & 11.97 &  8.68 &  6.67 &  3.30 \\ 
    \midrule
        \multirow{4}{*}{Prompt} 
        & BERT              &  2.52 &  1.69 & 26.74 & 29.02 \\ 
        & RoBERTa           & 12.62 & 10.51 & 21.01 & 20.26 \\ 
        & DistilBERT        &  8.67 &  7.55 & 22.18 & 27.34 \\ 
        & DistilRoBERTa     & 15.88 & 16.89 & 23.25 & 21.88 \\ 
    \midrule
        \multirow{4}{*}{TAG-Z}
        & BERT              & 13.07 & 11.45 & 37.16 & 41.46 \\
        & RoBERTa           & 35.86 & \textbf{37.22} & \textbf{47.08} & \textbf{53.20} \\
        & DistilBERT        & 28.96 & 28.29 & 35.86 & 39.61 \\
        & DistilRoBERTa     & \textbf{37.08} & 37.20 & 42.58 & 48.19 \\
    \bottomrule
    \end{tabular}
\end{table}

\begin{table*}[t!]
    \renewcommand{\arraystretch}{.9}
  \caption{Mean test ACC (\%) and F1 (\%) of TAG-F and the baselines with different numbers of training samples on \textsc{ogbn-arxiv}. The best scores are highlighted in bold. Those marked by $^{\star}$ are significantly better than those of the next-best scores with significance level $p<0.01$ based on the Mann-Whitney \emph{U} Test \cite{rank_sum_test}.}
  \label{tab:few-shot-arxiv}
  \centering
    \begin{tabular}{clcccccccccc}
    \toprule
    \multirow{2}{*}{\textbf{Input}} & \multirow{2}{*}{\textbf{Method}} & \multicolumn{2}{c}{\textbf{1-shot}}   & \multicolumn{2}{c}{\textbf{2-shot}} & \multicolumn{2}{c}{\textbf{3-shot}}  & \multicolumn{2}{c}{\textbf{5-shot}} & \multicolumn{2}{c}{\textbf{10-shot}}            \\
    \cmidrule(r){3-4} \cmidrule(r){5-6} \cmidrule(r){7-8} \cmidrule(r){9-10} \cmidrule(r){11-12}
          & & \textbf{ACC} & \textbf{F1} & \textbf{ACC} & \textbf{F1} & \textbf{ACC} & \textbf{F1} & \textbf{ACC} & \textbf{F1} & \textbf{ACC} & \textbf{F1} \\
    \midrule
    \multirow{7}{*}{OGB$^{\dagger}$}
    & MLP    &  5.28 &  5.28 &  9.94 & 10.69 &  9.82 & 10.31 & 14.68 & 16.01 & 17.91 & 20.27 \\ 
    & GCN    &  9.67 & 10.49 & 13.14 & 13.13 & 15.66 & 17.42 & 18.65 & 21.63 & 23.78 & 26.99 \\ 
    & GAT    & 11.36 & 11.63 & 13.69 & 15.35 & 19.99 & 22.25 & 23.29 & 26.78 & 28.67 & 32.66 \\ 
    & SGC    &  7.04 &  6.39 & 12.70 & 14.47 & 14.49 & 16.20 & 17.77 & 21.22 & 20.23 & 25.50 \\ 
    & GCNII  &  9.30 & 11.05 & 12.27 & 13.73 & 16.21 & 18.71 & 18.75 & 21.85 & 21.38 & 24.10 \\ 
    & UniMP  & 10.29 & 11.90 & 12.77 & 13.86 & 15.46 & 17.79 & 22.92 & 26.18 & 30.55 & 34.26 \\ 
    & RevGNN & 10.33 & 11.64 & 13.76 & 15.63 & 18.83 & 21.55 & 22.75 & 26.64 & 26.76 & 30.48 \\ 
    \midrule
    \multirow{7}{*}{SE}
    & MLP    &  9.40 & 10.61 & 11.93 & 13.00 & 14.95 & 16.50 & 20.65 & 22.21 & 27.57 & 30.09 \\ 
    & GCN    & 10.91 & 11.63 & 16.11 & 16.84 & 18.57 & 20.36 & 21.01 & 23.68 & 30.21 & 33.48 \\ 
    & GAT    & 12.90 & 14.73 & 15.17 & 16.57 & 22.11 & 24.78 & 27.00 & 30.32 & 34.67 & 38.62 \\ 
    & SGC    & 10.13 &  8.76 & 15.42 & 16.16 & 21.28 & 22.19 & 22.36 & 26.13 & 26.96 & 32.28 \\ 
    & GCNII  &  8.08 &  8.78 & 12.94 & 14.64 & 13.85 & 15.45 & 17.31 & 18.56 & 23.79 & 25.40 \\ 
    & UniMP  &  7.92 &  8.23 & 12.18 & 12.46 & 13.79 & 13.42 & 16.37 & 18.15 & 27.95 & 30.79 \\ 
    & RevGNN & 12.74 & 14.67 & 16.30 & 19.05 & 19.93 & 22.05 & 25.92 & 28.78 & 32.09 & 35.88 \\ 
    \midrule
    Raw Texts
    & TAG-F & \textbf{33.21}$^\star$ & \textbf{30.52}$^\star$ & \textbf{42.03}$^\star$ & \textbf{41.78}$^\star$ & \textbf{43.56}$^\star$ & \textbf{43.68}$^\star$ & \textbf{48.57}$^\star$ & \textbf{49.07}$^\star$ & \textbf{53.34}$^\star$ & \textbf{53.46}$^\star$ \\ 
    \bottomrule
  \end{tabular}
\end{table*}
\begin{table*}[t!]
    \renewcommand{\arraystretch}{.9}
  \caption{Mean test ACC (\%) and F1 (\%) of TAG-F and the baselines with different numbers of training samples on \textsc{ogbn-products}. The annotation criteria are the same as in Table \ref{tab:few-shot-arxiv}.}
  \label{tab:few-shot-products}
\centering
  \begin{tabular}{clcccccccccc}
    \toprule
    \multirow{2}{*}{\textbf{Input}} & \multirow{2}{*}{\textbf{Method}} & \multicolumn{2}{c}{\textbf{1-shot}}   & \multicolumn{2}{c}{\textbf{2-shot}} & \multicolumn{2}{c}{\textbf{3-shot}}  & \multicolumn{2}{c}{\textbf{5-shot}} & \multicolumn{2}{c}{\textbf{10-shot}}            \\
    \cmidrule(r){3-4} \cmidrule(r){5-6} \cmidrule(r){7-8} \cmidrule(r){9-10} \cmidrule(r){11-12}
          & & \textbf{ACC} & \textbf{F1} & \textbf{ACC} & \textbf{F1} & \textbf{ACC} & \textbf{F1} & \textbf{ACC} & \textbf{F1} & \textbf{ACC} & \textbf{F1} \\
    \midrule
    \multirow{7}{*}{OGB$^{\dagger}$}
    & MLP    &  4.12 &  5.41 &  5.66 &  7.14 &  6.76 &  8.81 &  9.12 & 11.65 & 11.37 & 14.05 \\ 
    & GCN    & 20.09 & 25.20 & 24.30 & 28.80 & 29.72 & 37.09 & 33.51 & 39.56 & 39.71 & 47.78 \\ 
    & GAT    &  6.89 &  8.54 &  7.65 &  9.09 & 13.65 & 16.96 & 19.65 & 23.44 & 29.89 & 36.22 \\ 
    & SGC    & 16.03 & 18.46 & 24.41 & 27.42 & 27.25 & 32.21 & 33.96 & 39.02 & 39.12 & 44.79 \\ 
    & GCNII  & 16.68 & 20.25 & 21.24 & 24.19 & 24.84 & 30.80 & 32.54 & 38.57 & 37.12 & 44.43 \\ 
    & UniMP  & 10.13 & 12.52 & 11.29 & 12.46 & 16.39 & 18.93 & 17.86 & 21.36 & 25.58 & 30.44 \\ 
    & RevGNN &  7.99 & 10.48 & 12.05 & 14.54 & 15.17 & 18.81 & 17.02 & 20.20 & 23.14 & 27.34 \\ 
    \midrule
    \multirow{7}{*}{SE}
    & MLP    & 11.77 & 15.32 & 12.62 & 15.54 & 15.46 & 19.20 & 20.46 & 24.43 & 30.00 & 36.02 \\ 
    & GCN    & 14.84 & 17.46 & 17.05 & 19.24 & 23.14 & 28.87 & 23.75 & 25.07 & 23.74 & 27.59 \\ 
    & GAT    & 23.71 & 28.48 & 23.42 & 28.26 & 36.86 & 44.12 & 40.55 & 47.89 & 51.62 & 58.50 \\ 
    & SGC    & 26.62 & 31.76 & 37.05 & 42.25 & 48.20 & 54.61 & 50.46 & 57.37 & 59.69 & 65.28 \\ 
    & GCNII  & 19.72 & 22.99 & 25.93 & 30.52 & 30.42 & 37.06 & 26.46 & 41.01 & 51.03 & 56.29 \\ 
    & UniMP  &  9.87 & 13.30 & 16.04 & 20.04 & 26.42 & 30.99 & 31.00 & 36.92 & 44.50 & 50.59 \\ 
    & RevGNN & 18.29 & 23.26 & 23.43 & 29.16 & 29.20 & 35.57 & 37.02 & 43.54 & 45.53 & 52.37 \\ 
    \midrule
    Raw Texts
    & TAG-F & \textbf{47.76}$^\star$ & \textbf{51.70}$^\star$ & \textbf{55.16}$^\star$ & \textbf{60.08}$^\star$ & \textbf{56.30}$^\star$  & \textbf{61.60}$^\star$ & \textbf{59.64}$^\star$ & \textbf{65.03}$^\star$ & \textbf{64.41}$^\star$ & \textbf{69.18}$^\star$ \\ 
    \bottomrule
  \end{tabular}
\end{table*}

To comprehensively understand how TAG performs in node classification with limited supervision, we investigate six different settings, which are the 0-, 1-, 2-, 3-, 5-, and 10-shot node classification tasks. Table \ref{tab:zero-shot-main} shows the results on the zero-shot case, while Table \ref{tab:few-shot-arxiv} and \ref{tab:few-shot-products} show the results on the few-shot cases.
The complete experiment results with standard deviation are shown in Table \ref{tab:few-shot-arxiv-appendix}, \ref{tab:few-shot-products-appendix} in Appendix section \ref{sec:app_exp}.

Overall, TAG significantly outperforms all the baselines in both datasets under all the settings. For the zero-shot setting, TAG-Z with RoBERTa surpasses the next-best performing baseline, which is Prompt, by 19.98\% ACC and 26.71\% F1 on \textsc{ogbn-arxiv}, and by 20.34\% ACC and 24.18\% F1 on \textsc{ogbn-products}. While equipped with other PMLMs, TAG also significantly improves the ACC and F1 of its Prompt counterparts. SES, regardless of which PMLM it uses, is inferior to prompt-based methods in most cases, and in some cases performs only slightly better than the three random naive baselines.

In the few-shot setting (Tables \ref{tab:few-shot-arxiv} and \ref{tab:few-shot-products}), TAG-F is consistently superior to the baselines, both those using OGB features, and those using SE features. For example, in the 2-shot setting, TAG-F outperforms the closest baseline by 25.73\% and 18.11\% ACC on \textsc{ogbn-arxiv} and \textsc{ogbn-products} respectively. Similar advantages are observed for the F1 metric. The performance discrepancy becomes smaller when more labeled samples are given (e.g., 10-shot), which is expected due to the powerful learning capacity of state-of-the-art GNN baselines. Interestingly, among these baselines, simple models such as GCN and SGC, instead, perform comparably or even better than recently proposed models such as UniMP and RevGNN. UniMP and RevGNN are powerful but more complex, which can lead to overfitting in the few-shot setting, despite their effectiveness in the standard semi-supervised learning setting. 

The statistical tests together with the above analyses provide evidence of large and statistically significant performance gains of TAG over competitive zero- and few-shot node classification approaches, and the benefits of multimodal learning in the text-and-graph setting.

\subsection{RQ2: Ablation Studies}
\subsubsection{Prompts used by TAG}
\begin{table*}[htbp]\small
    \renewcommand{\arraystretch}{.75}
  \caption{Mean test ACC (\%) and F1 (\%) of TAG on \textsc{ogbn-arxiv} and \textsc{ogbn-products} w.r.t. different instruction texts.}
  \label{tab:ablation-instruction-text}
\centering
  \begin{tabular}{cccccccccccccc}
    \toprule
    \multirow{4}{*}{\textbf{Dataset}} & \multirow{4}{*}{\textbf{Instruction text} \textit{t}} & 
    \multicolumn{2}{c}{\textbf{TAG-Z}} & \multicolumn{10}{c}{\textbf{TAG-F}} \\
    \cmidrule(r){3-4} \cmidrule(r){5-14}
    & & \multicolumn{2}{c}{\textbf{0-shot}} & \multicolumn{2}{c}{\textbf{1-shot}}   & \multicolumn{2}{c}{\textbf{2-shot}} & \multicolumn{2}{c}{\textbf{3-shot}}  & \multicolumn{2}{c}{\textbf{5-shot}} & \multicolumn{2}{c}{\textbf{10-shot}}            \\
    \cmidrule(r){3-4} \cmidrule(r){5-6} \cmidrule(r){7-8} \cmidrule(r){9-10} \cmidrule(r){11-12} \cmidrule(r){13-14}
    & & \textbf{ACC} & \textbf{F1} & \textbf{ACC} & \textbf{F1} & \textbf{ACC} & \textbf{F1} & \textbf{ACC} & \textbf{F1} & \textbf{ACC} & \textbf{F1} & \textbf{ACC} & \textbf{F1} \\
    \midrule
    \multirow{7}{*}{\rotatebox{90}{\textsc{ogbn-arxiv}}}
    & `'                                & 25.61 & 28.17 & 31.00 & 29.15 & 40.20 & 40.16 & 44.03 & 43.96 & 46.82 & 47.26 & 52.93 & 52.81 \\
    & `Topic'                           & 35.86 & 37.22 & 33.21 & 30.52 & 42.03 & 41.78 & 43.56 & 43.68 & 48.57 & 49.07 & 53.34 & 53.46 \\
    & `Category'                        & 36.97 & 37.97 & 32.18 & 29.17 & 43.15 & 42.72 & 45.20 & 45.25 & 48.58 & 48.72 & 52.48 & 53.51 \\
    & `Paper topic'                     & 33.50 & 35.04 & 32.04 & 28.86 & 41.52 & 40.74 & 43.21 & 43.90 & 48.36 & 49.04 & 53.16 & 53.54 \\
    & `Article category'                & 37.11 & 38.26 & 32.80 & 29.74 & 42.59 & 42.23 & 43.74 & 43.84 & 50.02 & 50.03 & 53.30 & 53.11 \\
    & `This paper talks about'          & 34.61 & 36.11 & 32.15 & 29.53 & 39.56 & 39.10 & 43.61 & 43.79 & 48.25 & 48.31 & 53.53 & 53.57 \\
    & `The following content is about'  & 35.33 & 36.70 & 32.60 & 29.44 & 43.11 & 42.66 & 45.03 & 45.09 & 49.46 & 49.45 & 54.37 & 54.34 \\
    \midrule
    \multirow{7}{*}{\rotatebox{90}{\textsc{ogbn-products}}}
    & `'                                        & 43.52 & 45.42 & 46.82 & 49.61 & 56.08 & 59.56 & 59.29 & 63.31 & 61.98 & 65.95 & 65.98 & 70.01 \\
    & `Category'                                & 44.68 & 48.94 & 47.20 & 50.78 & 56.47 & 60.20 & 59.48 & 63.44 & 63.32 & 67.51 & 64.83 & 69.15 \\
    & `Type'                                    & 42.18 & 47.82 & 46.29 & 49.96 & 54.61 & 59.31 & 56.28 & 61.50 & 57.84 & 63.07 & 64.72 & 69.22 \\
    & `Product Category'                        & 47.08 & 53.20 & 47.76 & 51.70 & 55.16 & 60.08 & 56.30 & 61.60 & 59.64 & 65.03 & 64.41 & 69.18 \\
    & `Item type'                               & 43.79 & 49.35 & 45.96 & 49.64 & 54.44 & 59.49 & 54.92 & 59.78 & 59.03 & 64.40 & 64.79 & 69.30 \\
    & `This product belongs to'                 & 46.10 & 50.09 & 44.20 & 47.62 & 55.73 & 59.05 & 59.89 & 63.69 & 61.83 & 66.31 & 65.93 & 70.21 \\
    & `The following product is categorized to' & 47.01 & 53.22 & 45.97 & 49.24 & 54.29 & 58.59 & 55.48 & 60.27 & 60.41 & 65.35 & 64.79 & 69.37 \\
    \bottomrule
  \end{tabular}
\end{table*}
As prompts form the foundation of TAG as a training-free prior knowledge retrieval tool, it is important to understand how different prompts, especially different instruction texts $t$ affect the node classification performance. Here, we additionally design six more instructions texts for each dataset and test them in both zero-shot (using TAG-Z) and few-shot (using TAG-F) settings. Table \ref{tab:ablation-instruction-text} shows the experimental results, from which we have the following observations:
\begin{itemize}
    \item TAG is generally robust to different instruction texts. Compared to our default instruction text selections (`Topic' for \textsc{ogbn-arxiv} and 'Product Category' for \textsc{ogbn-products}), TAG with these new instruction texts yields similar zero- and few-shot results in most cases, while still outperforming all the zero- and few-shot baselines. This suggests that the choice of instruction texts does not impact TAG's overall performance significantly, which makes TAG a more flexible zero- and few-shot node classification model in practice. 
    \item Task-relevant instruction texts work better than empty instructions, especially in zero-shot setting. For instance, in \textsc{ogbn-arxiv}, the zero-shot ACC is only 25.61\%, which is around 10\% lower than the other variants with relevant instruction texts. Similarly, TAG-Z with empty instruction performs the worst in zero-shot F1 (second worst in zero-shot ACC) among all TAG-Z models with other instruction texts. The performance gaps between empty instructions and task-relevant instructions, however, become narrower when more training samples are given. 
\end{itemize}

\subsubsection{Components of TAG}
We further introduce several ablated variants of TAG by removing some of the components to study the necessity of their inclusion. For TAG-Z, we remove the normalization (-\textsc{Norm}) and/or the graph propagation (-\textsc{GP}). For TAG-F, we keep the PLG module complete, but remove the identity connection in Eq. (\ref{eq:id}) (-\textsc{ID}) and/or the shrinkage ensemble (-\textsc{Ens}) (i.e., using only one non-shrunk trainable model). Table \ref{tab:ablation-components} summarizes the experimental results providing us with the following findings:
\begin{table}[htbp]
    \renewcommand{\arraystretch}{.9}
    \caption{Experimental results of TAG and its ablated variants under different few-shot settings.}
    \label{tab:ablation-components}
    \centering
    \begin{tabular}{clcccc}
    \toprule
        & & \multicolumn{2}{c}{\textsc{arxiv}} & \multicolumn{2}{c}{\textsc{products}} \\
        \cmidrule(r){3-4} \cmidrule(r){5-6}
        \textbf{Setting} &  \textbf{Method} & \textbf{ACC} & \textbf{F1} & \textbf{ACC} & \textbf{F1}\\
        \midrule
        \multirow{5}{*}{0-shot} 
        & TAG-Z                                         & \textbf{35.79} & \textbf{37.38} & \textbf{47.08} & \textbf{53.20} \\
        \cmidrule(l){2-6}
        & -\textsc{Norm}                                & 10.01 &  3.65 & 18.76 & 16.94 \\
        & -\textsc{GP}                                  & 15.17 & 16.99 & 25.75 & 32.58 \\
        & -\textsc{Norm} \& -\textsc{GP}                & 12.62 & 10.51 & 21.01 & 20.26 \\
        \midrule
        \multirow{5}{*}{3-shot} 
        & TAG-F                                         & \textbf{43.56} & \textbf{43.68} & \textbf{56.30}  & \textbf{61.60} \\
        \cmidrule(l){2-6}
        & -\textsc{ID}                                  & 38.38 & 38.63 & 43.35 & 49.08 \\
        & -\textsc{Ens}                                 & 38.29 & 38.24 & 44.91 & 50.25 \\
        & -\textsc{ID} \& -\textsc{Ens}                 & 32.30 & 32.94 & 30.22 & 34.14 \\
        \midrule
        \multirow{5}{*}{5-shot} 
        & TAG-F                                         & \textbf{48.57} & \textbf{49.07} & \textbf{59.64} & \textbf{65.03} \\
        \cmidrule(l){2-6}
        & -\textsc{ID}                                  & 41.80 & 41.72 & 48.22 & 52.89 \\
        & -\textsc{Ens}                                 & 41.40 & 41.99 & 49.35 & 53.57 \\
        & -\textsc{ID} \& -\textsc{Ens}                 & 35.14 & 35.18 & 40.94 & 43.56 \\
    \bottomrule
    \end{tabular}
\end{table}
\begin{itemize}
    \item Both graph propagation and normalization are important to generate high-quality prior logits. The lack of normalization in PLG causes the largest performance drops in both datasets. This is due to the biased distributions of the score matrices in both datasets, which can be solved by normalizing these matrices. 
    \item The identity connection (prior logits), despite not affecting the expressiveness of TAG-F, greatly improves few-shot learning. Without the identity connection, the classification accuracy decreases, for instance, by 6.77\% under 5-shot setting in \textsc{ogbn-arxiv}. The performance gaps are more obvious when fewer training samples are given in \textsc{ogbn-products}. 
    \item Shrinkage ensemble further improves the classification performance of TAG-F. Since few-shot training leads to high variability, the ensemble models benefit from shrinkage toward the prior and the the multiple predictions from individual models.
\end{itemize}
The above findings, in brief, suggest that all these components play essential roles in improving the zero- and few-shot node classification performance of TAG.

\subsection{RQ3: Effects of Hyperparameters}

\begin{figure}[t]\tiny
    \centering
    \includegraphics[width=3.3in]{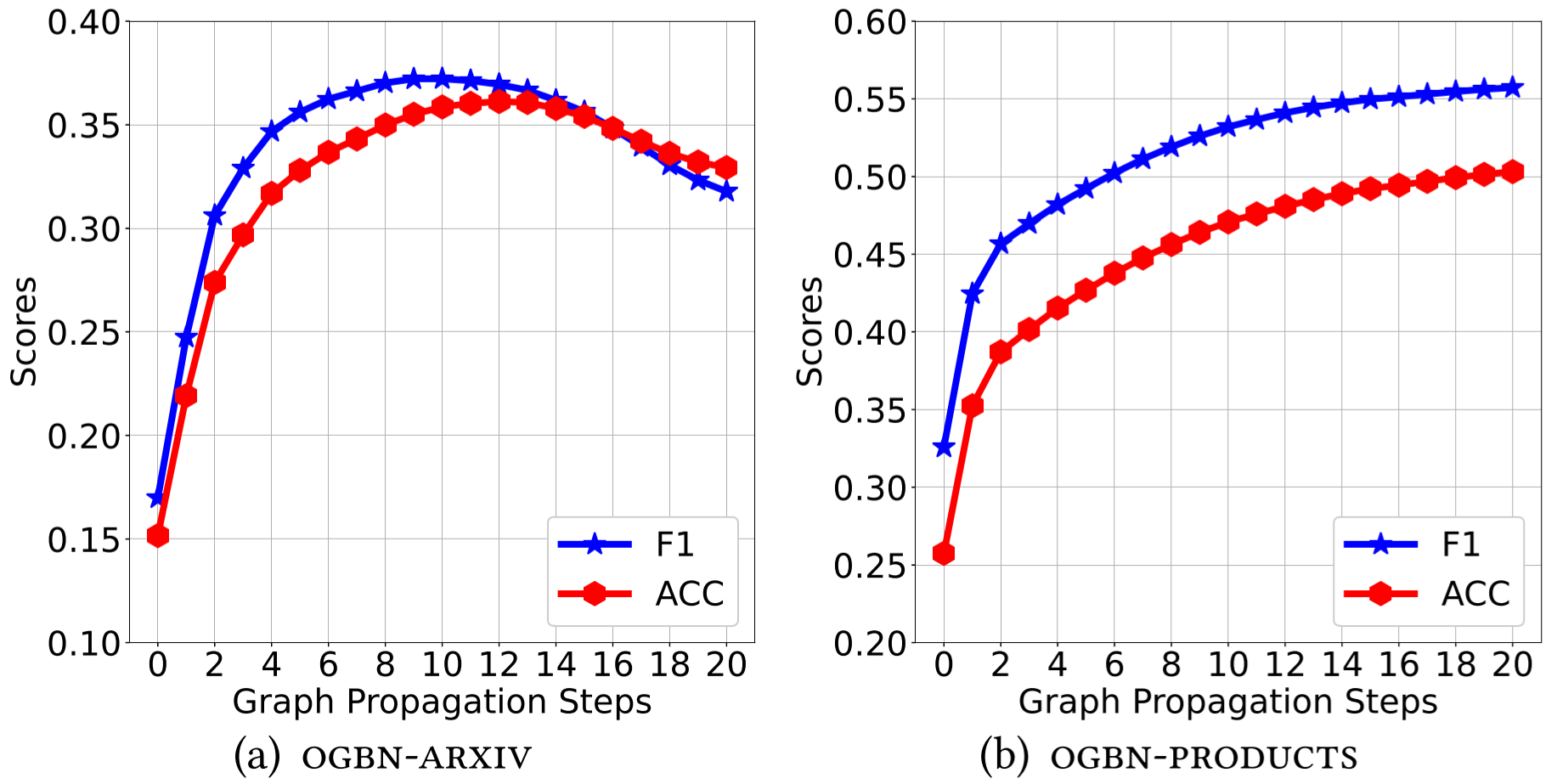}
    \caption{Zero-shot test ACC and F1 of TAG-Z w.r.t different graph propagation steps $P$ on the two datasets.}
    \label{exp:graph_propagation}
\end{figure}

\begin{figure}[t]\tiny
    \centering
    \includegraphics[width=3.3in]{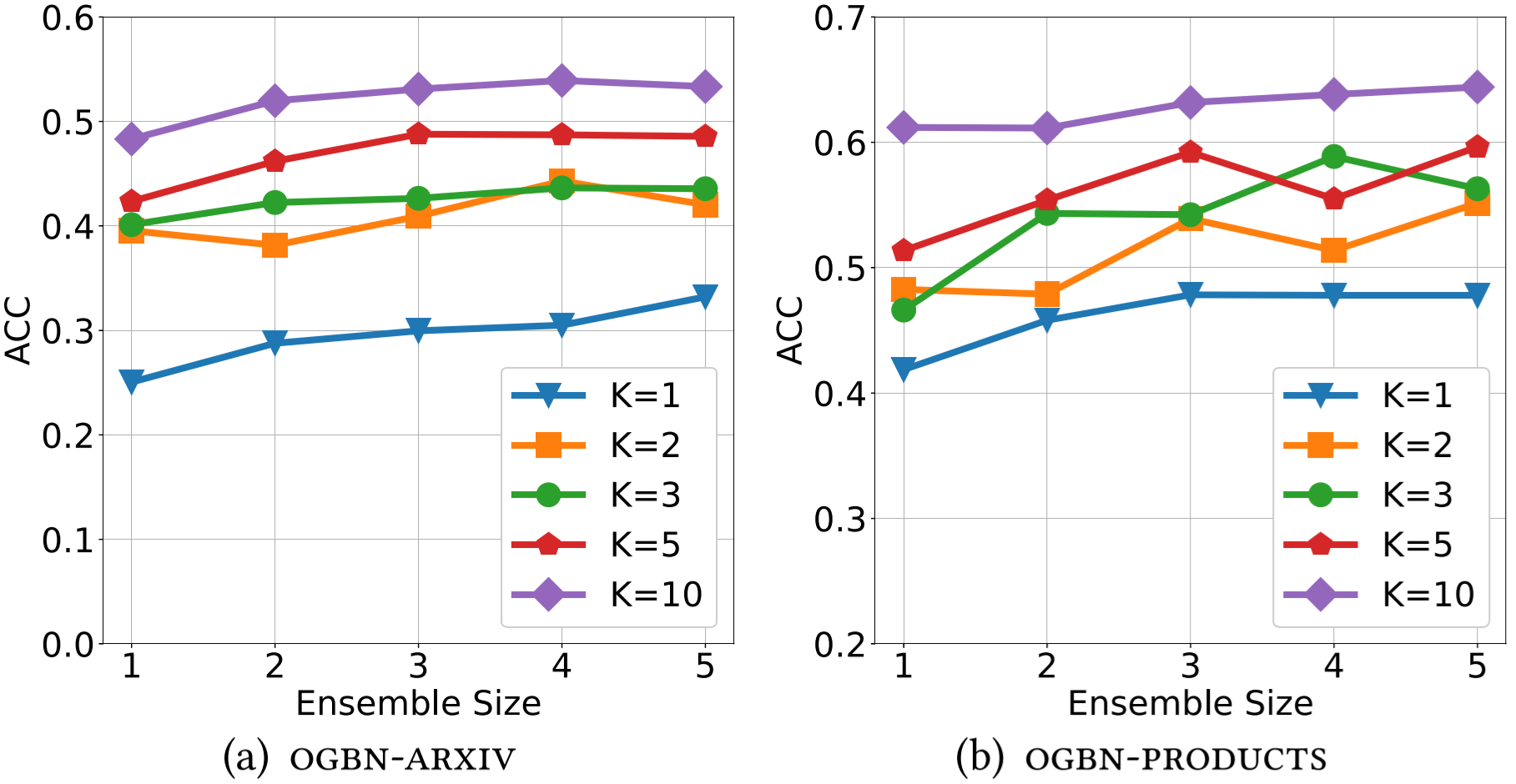}
    \caption{Mean test ACC of TAG-F w.r.t. different ensemble size $n_e$ on the two datasets.}
    \label{exp:ensemble_size}
\end{figure}

\begin{figure}[t]\tiny
    \centering
    \includegraphics[width=3.3in]{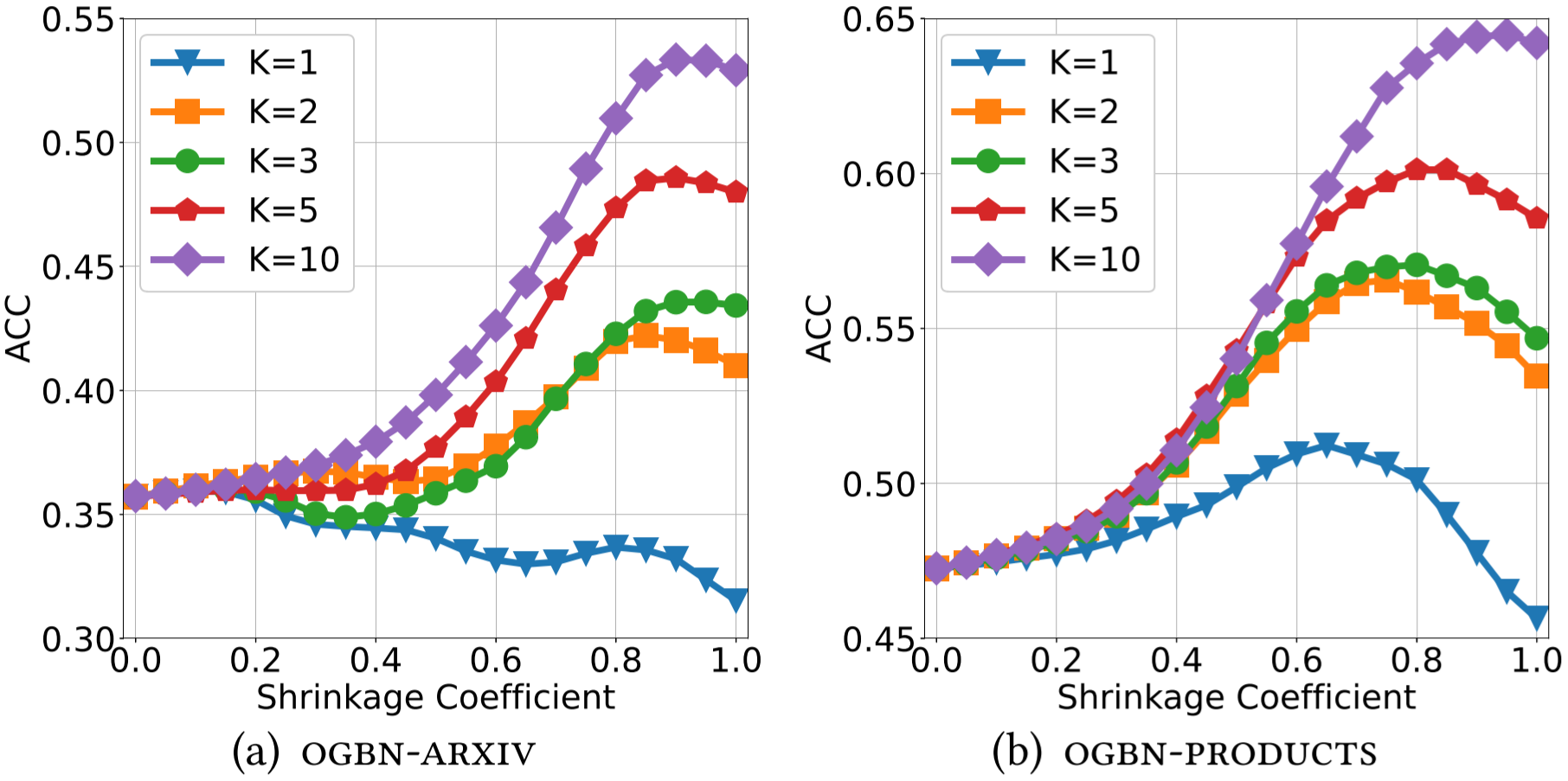}
    \caption{Mean test ACC of TAG-F w.r.t. different shrinkage coefficient $\alpha$ on the two datasets.}
    \label{exp:shrinkage}
\end{figure}

In this section, we investigate the effects of TAG's hyperparameters on its zero- and few-shot node classification performance. We consider three hyperparameters: the graph propagation steps $P$, the ensemble size $n_e$, and the shrinkage coefficient $\alpha$, in the following experiments. We vary one parameter at a time, while keeping the remaining parts of TAG unchanged.
\begin{itemize}

    \item \textbf{Graph Propagation Steps $P$}: Figure \ref{exp:graph_propagation} shows that introducing graph propagation clearly improves the results (compared to $P=0$). Within a reasonable range between $5$ and $20$, performance is relatively stable and high throughout, clearly outperforming the baseline approaches throughout this range, so we recommend $10$ as a default choice.

    \item \textbf{Ensemble Size $n_e$}: As shown in Figure \ref{exp:ensemble_size}, a small ensemble size (e.g., $n_e=3$) already leads to distinguishable performance gains in both datasets, while further increasing $n_e$ improves these results much more mildly. Since larger ensembles are generally expected to be beneficial overall, we recommend $n_e=5$.

    \item \textbf{Shrinkage Coefficient $\alpha$}: Figure \ref{exp:shrinkage} shows that introducing shrinkage (i.e., decreasing $\alpha$ from $1$ to values below $1$) improves the accuracy of TAG-F in both datasets under all few-shot settings. This validates the motivation of TAG-F, that relying more on the prior logits can improve the classification performance in few-shot settings. More extreme shrinkage values (e.g., $0.5$ or below) cause excessive changes to the model and may degrade performance, but within a mild range of $(0.8, 1)$, performance is overall high and relatively stable, so we recommend a value in this range, with $0.9$ as a safe default.
\end{itemize}


To sum up, the introduction of these hyperparameters all lead to improvements in TAG's node classification performance, and its performance is stable and high throughout over reasonable ranges.

\section{Conclusion}
In this paper, we explored the benefits of a multimodal approach in the text and graph setting, under the zero- and few-shot settings. Instead of starting with precomputed text features like many existing approaches, we use raw texts as input, integrated into the model design in combination with graph representation learning. Specifically, we introduce prompts into our proposed approach TAG so that we directly acquire prior knowledge from PMLMs for zero- and few-shot node classification, rather than  performing meta-training in meta-learning-based approaches. As a result, TAG effectively learns from limited supervised signals without complex meta-learning procedures. Experiments demonstrate that TAG excels at zero- and few-shot node classification and surpasses baselines by large and statistically significant margins. Going further, an interesting question is about the extent to which the benefits of designing deeply multimodal approaches can also extend to other graph-related tasks (e.g., link and graph classification) as well as to cases where labelled data is relatively abundant.


\bibliographystyle{ACM-Reference-Format}
\bibliography{kdd}

\appendix
\section{Experimental Setup}\label{sec:app_setup}


\begin{table}[htbp]
\renewcommand{\arraystretch}{.75}
    \caption{Details on the hyperparameter configurations of the few-shot baselines.}
    \label{app:hyper_appendix}
    \centering
    \begin{tabular}{lcc}
    \toprule
        \textbf{Method} & \textbf{Hidden Channels} & \textbf{Remarks} \\
    \midrule
        MLP     & 128   & NA\\
        GCN     & 128   & NA \\
        GAT     & 32    & heads=4 \\
        SGC     & NA    & only one layer \\
        GCNII   & 128   & NA \\
        UniMP   & 64   & label rate=0.625, head=2\\
        RevGNN  & 64    & groups=2\\
    \bottomrule
    \end{tabular}
\end{table}



\section{Experimental Results}\label{sec:app_exp}
\subsection{Few-shot Experimental Results with Standard Deviation}
\begin{table*}[htbp]\small
    \renewcommand{\arraystretch}{.75}
  \caption{Mean test ACC (\%) and F1 (\%) and their standard deviation (in parenthesis) over 5 repeats of TAG-F and the baselines with different numbers of training samples on \textsc{ogbn-arxiv}. The annotation criteria are the same as in Table \ref{tab:few-shot-arxiv}.}
  \label{tab:few-shot-arxiv-appendix}
  \centering
    \begin{tabular}{clcccccccccc}
    \toprule
    \multirow{2}{*}{\textbf{Input}} & \multirow{2}{*}{\textbf{Method}} & \multicolumn{2}{c}{\textbf{1-shot}}   & \multicolumn{2}{c}{\textbf{2-shot}} & \multicolumn{2}{c}{\textbf{3-shot}}  & \multicolumn{2}{c}{\textbf{5-shot}} & \multicolumn{2}{c}{\textbf{10-shot}}            \\
    \cmidrule(r){3-4} \cmidrule(r){5-6} \cmidrule(r){7-8} \cmidrule(r){9-10} \cmidrule(r){11-12}
          & & \textbf{ACC} & \textbf{F1} & \textbf{ACC} & \textbf{F1} & \textbf{ACC} & \textbf{F1} & \textbf{ACC} & \textbf{F1} & \textbf{ACC} & \textbf{F1} \\
    \midrule
    \multirow{14}{*}{OGB$^{\dagger}$}
    & MLP    &  5.28 &  5.28 &  9.94 & 10.69 &  9.82 & 10.31 & 14.68 & 16.01 & 17.91 & 20.27 \\ 
    &        & (1.34)& (1.70)& (1.59)& (1.90)& (3.39)& (3.43)& (2.17)& (2.34)& (1.74)& (1.87)\\
    & GCN    &  9.67 & 10.49 & 13.14 & 13.13 & 15.66 & 17.42 & 18.65 & 21.63 & 23.78 & 26.99 \\ 
    &        & (1.90)& (1.35)& (3.79)& (3.34)& (2.33)& (2.48)& (0.68)& (0.69)& (1.27)& (1.58)\\
    & GAT    & 11.36 & 11.63 & 13.69 & 15.35 & 19.99 & 22.25 & 23.29 & 26.78 & 28.67 & 32.66 \\ 
    &        & (1.52)& (2.21)& (2.64)& (2.61)& (3.77)& (3.68)& (1.63)& (1.74)& (1.90)& (1.97)\\
    & SGC    &  7.04 &  6.39 & 12.70 & 14.47 & 14.49 & 16.20 & 17.77 & 21.22 & 20.23 & 25.50 \\ 
    &        & (1.20)& (1.48)& (2.15)& (2.42)& (1.31)& (2.18)& (1.16)& (1.40)& (2.21)& (2.59)\\
    & GCNII  &  9.30 & 11.05 & 12.27 & 13.73 & 16.21 & 18.71 & 18.75 & 21.85 & 21.38 & 24.10 \\ 
    &        & (1.63)& (2.11)& (2.40)& (2.95)& (2.66)& (3.39)& (4.34)& (5.12)& (2.26)& (2.57)\\
    & UniMP  & 10.29 & 11.90 & 12.77 & 13.86 & 15.46 & 17.79 & 22.92 & 26.18 & 30.55 & 34.26 \\ 
    &        & (1.99)& (2.84)& (1.46)& (1.78)& (4.92)& (5.57)& (3.89)& (4.07)& (3.45)& (3.31)\\
    & RevGNN & 10.33 & 11.64 & 13.76 & 15.63 & 18.83 & 21.55 & 22.75 & 26.64 & 26.76 & 30.48 \\ 
    &        & (3.37)& (4.16)& (1.43)& (1.84)& (2.14)& (2.33)& (1.62)& (1.73)& (1.47)& (1.60)\\
    \midrule
    \multirow{14}{*}{SE}
    & MLP    &  9.40 & 10.61 & 11.93 & 13.00 & 14.95 & 16.50 & 20.65 & 22.21 & 27.57 & 30.09 \\ 
    &        & (1.94)& (1.88)& (1.62)& (2.14)& (1.78)& (1.68)& (1.84)& (1.57)& (0.75)& (1.62)\\
    & GCN    & 10.91 & 11.63 & 16.11 & 16.84 & 18.57 & 20.36 & 21.01 & 23.68 & 30.21 & 33.48 \\ 
    &        & (3.46)& (3.29)& (3.46)& (3.90)& (2.77)& (2.94)& (2.15)& (2.67)& (2.43)& (2.32)\\
    & GAT    & 12.90 & 14.73 & 15.17 & 16.57 & 22.11 & 24.78 & 27.00 & 30.32 & 34.67 & 38.62 \\ 
    &        & (3.01)& (2.93)& (2.68)& (2.95)& (0.94)& (0.83)& (0.97)& (0.71)& (1.19)& (1.45)\\
    & SGC    & 10.13 &  8.76 & 15.42 & 16.16 & 21.28 & 22.19 & 22.36 & 26.13 & 26.96 & 32.28 \\ 
    &        & (2.32)& (1.79)& (3.29)& (3.68)& (1.47)& (1.37)& (1.26)& (1.75)& (2.61)& (2.88)\\
    & GCNII  &  8.08 &  8.78 & 12.94 & 14.64 & 13.85 & 15.45 & 17.31 & 18.56 & 23.79 & 25.40 \\ 
    &        & (1.24)& (1.76)& (2.45)& (3.20)& (2.32)& (3.22)& (2.08)& (2.90)& (5.20)& (4.69)\\
    & UniMP  &  7.92 &  8.23 & 12.18 & 12.46 & 13.79 & 13.42 & 16.37 & 18.15 & 27.95 & 30.79 \\ 
    &        & (3.60)& (3.36)& (2.56)& (2.38)& (3.48)& (4.02)& (4.71)& (5.87)& (1.76)& (2.59)\\
    & RevGNN & 12.74 & 14.67 & 16.30 & 19.05 & 19.93 & 22.05 & 25.92 & 28.78 & 32.09 & 35.88 \\ 
    &        & (2.91)& (3.26)& (2.20)& (2.34)& (0.86)& (1.29)& (2.21)& (2.51)& (1.18)& (1.20)\\
    \midrule
    Raw Texts
    & TAG-F & \textbf{33.21}$^\star$ & \textbf{30.52}$^\star$ & \textbf{42.03}$^\star$ & \textbf{41.78}$^\star$ & \textbf{43.56}$^\star$ & \textbf{43.68}$^\star$ & \textbf{48.57}$^\star$ & \textbf{49.07}$^\star$ & \textbf{53.34}$^\star$ & \textbf{53.46}$^\star$ \\ 
    &       & (7.74)& (6.75)& (5.20)& (5.03)& (4.88)& (5.31)& (3.46)& (2.85)& (0.92)& (0.89) \\
    \bottomrule
  \end{tabular}
\end{table*}
\begin{table*}[htbp]\small
    \renewcommand{\arraystretch}{.75}
  \caption{Mean test ACC (\%) and F1 (\%) and their standard deviation (in parenthesis) over 5 repeats of TAG-F and the baselines with different numbers of training samples on \textsc{ogbn-products}. The annotation criteria are the same as in Table \ref{tab:few-shot-arxiv}.}
  \label{tab:few-shot-products-appendix}
\centering
  \begin{tabular}{clcccccccccc}
    \toprule
    \multirow{2}{*}{\textbf{Input}} & \multirow{2}{*}{\textbf{Method}} & \multicolumn{2}{c}{\textbf{1-shot}}   & \multicolumn{2}{c}{\textbf{2-shot}} & \multicolumn{2}{c}{\textbf{3-shot}}  & \multicolumn{2}{c}{\textbf{5-shot}} & \multicolumn{2}{c}{\textbf{10-shot}}            \\
    \cmidrule(r){3-4} \cmidrule(r){5-6} \cmidrule(r){7-8} \cmidrule(r){9-10} \cmidrule(r){11-12}
          & & \textbf{ACC} & \textbf{F1} & \textbf{ACC} & \textbf{F1} & \textbf{ACC} & \textbf{F1} & \textbf{ACC} & \textbf{F1} & \textbf{ACC} & \textbf{F1} \\
    \midrule
    \multirow{14}{*}{OGB$^{\dagger}$}
    & MLP    &  4.12 &  5.41 &  5.66 &  7.14 &  6.76 &  8.81 &  9.12 & 11.65 & 11.37 & 14.05 \\ 
    &        & (0.51)& (0.54)& (0.78)& (1.29)& (1.14)& (1.67)& (1.31)& (2.08)& (0.61)& (1.22)\\
    & GCN    & 20.09 & 25.20 & 24.30 & 28.80 & 29.72 & 37.09 & 33.51 & 39.56 & 39.71 & 47.78 \\ 
    &        & (3.19)& (4.61)& (2.41)& (3.46)& (0.74)& (0.63)& (2.38)& (4.03)& (2.91)& (2.89)\\
    & GAT    &  6.89 &  8.54 &  7.65 &  9.09 & 13.65 & 16.96 & 19.65 & 23.44 & 29.89 & 36.22 \\ 
    &        & (2.19)& (3.33)& (2.35)& (2.88)& (3.96)& (4.59)& (3.10)& (3.90)& (2.92)& (3.33)\\
    & SGC    & 16.03 & 18.46 & 24.41 & 27.42 & 27.25 & 32.21 & 33.96 & 39.02 & 39.12 & 44.79 \\ 
    &        & (1.46)& (1.68)& (1.91)& (3.40)& (2.19)& (3.26)& (3.09)& (5.23)& (1.73)& (2.58)\\
    & GCNII  & 16.68 & 20.25 & 21.24 & 24.19 & 24.84 & 30.80 & 32.54 & 38.57 & 37.12 & 44.43 \\ 
    &        & (2.75)& (4.13)& (5.60)& (6.95)& (1.64)& (2.13)& (3.83)& (5.21)& (3.06)& (3.55)\\
    & UniMP  & 10.13 & 12.52 & 11.29 & 12.46 & 16.39 & 18.93 & 17.86 & 21.36 & 25.58 & 30.44 \\ 
    &        & (1.94)& (2.04)& (1.56)& (1.93)& (2.11)& (2.26)& (2.46)& (2.77)& (1.71)& (1.80)\\
    & RevGNN &  7.99 & 10.48 & 12.05 & 14.54 & 15.17 & 18.81 & 17.02 & 20.20 & 23.14 & 27.34 \\ 
    &        & (0.94)& (1.34)& (1.13)& (1.87)& (1.08)& (1.72)& (1.75)& (2.17)& (0.81)& (1.24)\\
    \midrule
    \multirow{14}{*}{SE}
    & MLP    & 11.77 & 15.32 & 12.62 & 15.54 & 15.46 & 19.20 & 20.46 & 24.43 & 30.00 & 36.02 \\ 
    &        & (1.60)& (1.69)& (2.92)& (3.49)& (1.58)& (1.77)& (2.97)& (3.18)& (2.04)& (2.01)\\
    & GCN    & 14.84 & 17.46 & 17.05 & 19.24 & 23.14 & 28.87 & 23.75 & 25.07 & 23.74 & 27.59 \\ 
    &        & (3.25)& (5.52)& (5.18)& (5.19)& (6.45)& (8.42)& (5.19)& (6.04)& (7.24)& (7.72)\\
    & GAT    & 23.71 & 28.48 & 23.42 & 28.26 & 36.86 & 44.12 & 40.55 & 47.89 & 51.62 & 58.50 \\ 
    &        & (5.27)& (7.19)& (4.15)& (5.39)& (4.16)& (4.03)& (3.01)& (3.43)& (2.66)& (2.00)\\
    & SGC    & 26.62 & 31.76 & 37.05 & 42.25 & 48.20 & 54.61 & 50.46 & 57.37 & 59.69 & 65.28 \\ 
    &        & (3.87)& (5.94)& (5.36)& (6.62)& (3.44)& (3.62)& (2.13)& (2.21)& (0.41)& (0.42)\\
    & GCNII  & 19.72 & 22.99 & 25.93 & 30.52 & 30.42 & 37.06 & 26.46 & 41.01 & 51.03 & 56.29 \\ 
    &        & (6.40)& (7.74)& (2.95)& (3.47)& (6.57)& (7.17)& (7.21)& (9.43)& (2.77)& (2.47)\\
    & UniMP  &  9.87 & 13.30 & 16.04 & 20.04 & 26.42 & 30.99 & 31.00 & 36.92 & 44.50 & 50.59 \\ 
    &        & (3.32)& (4.83)& (2.58)& (3.08)& (8.74)&(10.01)& (5.13)& (5.90)& (6.58)& (7.04)\\
    & RevGNN & 18.29 & 23.26 & 23.43 & 29.16 & 29.20 & 35.57 & 37.02 & 43.54 & 45.53 & 52.37 \\ 
    &        & (3.19)& (4.16)& (3.45)& (3.48)& (2.84)& (4.27)& (2.99)& (3.11)& (2.51)& (2.26)\\
    \midrule
    Raw Texts
    & TAG-F & \textbf{47.76}$^\star$ & \textbf{51.70}$^\star$ & \textbf{55.16}$^\star$ & \textbf{60.08}$^\star$ & \textbf{56.30}$^\star$  & \textbf{61.60}$^\star$ & \textbf{59.64}$^\star$ & \textbf{65.03}$^\star$ & \textbf{64.41}$^\star$ & \textbf{69.18}$^\star$ \\ 
    &       &( 2.55)&( 2.44)&( 3.76)&( 2.92)&( 3.24)&( 2.39)&( 0.64)&( 0.99)&( 1.25)&( 1.14) \\
    \bottomrule
  \end{tabular}
\end{table*}

\end{document}